\documentclass[%
 reprint,
showpacs,
 amsmath,amssymb,
 aps,
 onecolumn,
 notitlepage,
 pre
]{revtex4-1}

\usepackage{graphicx}
\usepackage{dcolumn}
\usepackage{bm}
\usepackage{color}
\usepackage{ulem}

\begin{document}

\preprint{APS/123-QED}

\title{Three-wave interactions among surface gravity waves in a cylindrical container}

\author{Guillaume Michel}
\email{guillaume.michel@ens.fr}
 \affiliation{%
 Laboratoire de Physique Statistique, \'Ecole Normale Sup\'erieure, PSL Research University, Universit\'e Paris Diderot Sorbonne Paris-Cit\'e, Sorbonne Universit\'es UPMC, Univ. Paris 06 and CNRS, 24 Rue Lhomond, 75005 Paris, France}
\date{\today}
\begin{abstract}
The motion of a container filled with fluid perturbs the free surface and may result in spilling. In practice, most of the energy is localized in the modes of lowest frequencies (the gravest modes), and sloshing can be predicted once the dynamics of these modes is known. In this Rapid Communication, we investigate the nonlinear interactions between such grave modes in a cylindrical container. We first show that energy can be transferred from modes to modes with three-wave interactions: we derive the resonance conditions and characterize the early stage of this interaction. This result strongly contrasts with resonant interactions between surface gravity waves in extended domains such as the ocean, which involve at least four waves and are thus less efficient. An experiment is then performed to provide evidence of these nonlinear interactions.
\end{abstract}

\maketitle

\textit{Introduction.} Although the study of sloshing is almost two centuries old \cite{Poisson1828}, it is still extremely challenging to predict the surface deformation of a container undergoing back-and-forth oscillations. The main source of difficulties is nonlinearities, which must be taken into account starting from very small amplitudes: The experiments of Cocciaro \textit{et al.} \cite{Cocciaro1991,Cocciaro1993} have proved that the first nonlinearity to occur corrects the dissipation, and it has been recently demonstrated that, in perfect wetting, this effect becomes sizable when the sloshing amplitude compares to the thickness of the boundary layers (a fraction of millimetre in everyday life containers) \cite{Michel2017}. This explains why, whereas natural frequencies are experimentally found in good agreement with linear sloshing theory \cite{Cocciaro1991,Cocciaro1993, Case1956,Henderson1994, Howell2000}, damping rates involve larger discrepancies \cite{Case1956,Cocciaro1991,Henderson1994,Howell2000}. 

As the forcing amplitude is further increased, other nonlinear phenomena arise. The natural frequencies evolve with the oscillation amplitude \cite{Mack1962, Fultz1963}, and spatial and temporal harmonics appear, referred to as \textit{bound waves} in oceanography and first studied by Stokes for plane waves \cite{Stokes}. The dynamics of the different modes get coupled by nonlinearities: for a cylindrical container, back-and-forth displacement at the gravest eigenfrequency may, for instance, result in a pattern rotating with a periodic or a chaotic dynamics \cite{Miles1984a,Miles1984b}. Energy can also be transferred between modes of different eigenfrequencies by resonant interactions. Such energy transfers have been mainly studied in the context of oceanography, i.e., for surface gravity waves in a fluid of infinite depth and without lateral boundaries. In that case, resonant interactions involve at least four waves \cite{Phillips1960}, although three-wave interactions have been recently reported in the presence of a turbulent wave background \cite{Aubourg2015, Aubourg2016}. The associated growth rate and phase locking have been first computed by Longuet-Higgins \cite{LH1962}, and accurately describe experimental results for a large range of parameters (see \cite{Bonnefoy2016} and references therein). 

In this Rapid Communication, we investigate how these resonant energy transfers are affected by the presence of solid boundaries. We first consider two mother waves in a resonant configuration and compute the characteristics of the daughter wave for the early stage of the interaction: three-wave resonant interactions are demonstrated to be allowed in a cylindrical geometry. We explicitly compute the growth rate as a function of the different parameters. Then, we report on an experiment able to isolate a single three-wave resonant interaction in the absence of a wave-turbulent background. The characteristics of the daughter wave are found in agreement with the theoretical predictions.

\textit{Characteristics of the resonant interactions.} Consider a cylindrical container of radius $R$ partially filled with a fluid of density $\rho$ and of surface tension $\sigma$ up to a height $h$. This height is thereafter assumed to be infinite, and surface tension is disregarded. In the linear potential theory of surface waves, the surface elevation $\eta(r,\theta ,t)$ can then be decomposed into a sum of modes progressing clockwise or counter clockwise of the form \cite{Ibrahim}
\begin{equation}
A_{n,m} J_m ( k_{n,m} r) \mathrm{cos}(\omega_{n,m} t - m \theta),\label{surface_elevation}
\end{equation} 
where $A_{n,m}$ is the amplitude of the mode, $J_m$ is the Bessel function of order $m$ ($m$ is an integer that can be negative), $k_{n,m}$ is the wave number and is such that $k_{n,m}R$ is the $n$th root of $J_m'$, and $\omega_{n,m}=\sqrt{gk_{n,m}}$ is the angular frequency. In the linear and inviscid theory, the amplitudes $A_{n,m}$ are constant. However, if nonlinearities are taken into account, these coefficients become coupled and time-dependent, which may result in resonant energy transfers. For plane waves, these resonant interactions have been characterized either by an expansion of the continuity equation and boundary conditions at the free surface \cite{LH1962, Hasselmann_a},  or through a Hamiltonian formulation of the problem \cite{zakharov, Janssen}. Let $A$ and $B$ denote two waves, thereafter called mother waves, assumed to be of constant amplitudes and of surface elevations
\begin{equation}
\eta_i(r,\theta,t) = \frac{\epsilon_i}{k_i} J_{m_i} (k_i r) \mathrm{cos}(\omega_i  t - m_i \theta)  , \label{eta_i}
\end{equation}
where $i=\{A,B\}$ and $\epsilon_i$ is the steepness, i.e., a characteristic slope of the interface. The following result is derived in the appendix: three-wave resonant interactions occur if $J_{m_C}'(\omega_C^2 R/ g)=0$, with $\omega_C = \omega_A \pm \omega_B$ and $m_C = m_A \pm m_B$ (the signs have to be the same in these two equations), i.e., if a free wave of angular frequency $\omega_C$ and index $m_C$ exists. In that case, this daughter wave grows according to 
\begin{equation}
k_C \eta_C(r,\theta,t) = - \kappa \frac{\omega_C ^3}{4 \omega_A \omega_B} \epsilon_A \epsilon_B t J_{m_C}(k_C r) \mathrm{sin}(\omega_C t - m_C \theta), \label{DW}
\end{equation}
where $k_C = \omega_C^2 /g$ is the wave number and $\kappa$ is a dimensionless constant reported in eq. \eqref{kappa}.

The difference between resonant interactions in a cylindrical container and in the ocean, which respectively involve at least three and four waves, can be traced back to conservation laws. Indeed, both interactions are associated with conserved quantities, as in particular the energy $E$, the linear pseudo-momentum $\mathbf{P}$ and the angular pseudo-momentum along the vertical axis $L_z$. For a surface wave of the form \eqref{eta_i} of small steepness in a cylinder, the linear pseudo-momentum vanishes whereas the angular pseudo-momentum is given by $L_z = m E/\omega$: the resonance conditions $\omega_C = \omega_A \pm \omega_B$ and $m_C = m_A \pm m_B$ therefore respectively correspond to the conservation of energy and of angular pseudo-momentum. In contrast, plane progressive waves possess no angular pseudo-momentum, but possess a linear pseudo-momentum $\mathbf{P} = \mathbf{k}E/\omega$, and the resonant conditions become  $\omega_C = \omega_A \pm \omega_B$ and $\mathbf{k}_C = \mathbf{k}_A \pm \mathbf{k}_B$. Both these resonant conditions are supplemented with the dispersion relation for the wave $C$, i.e., require the daughter wave to be a free wave: $\omega_C^2 = g k_C$, with either $k_C = \xi_{m_C,n_C}/R$ or $k_C = \vert \mathbf{k}_C \vert$. The main effect of solid boundaries is that, whereas this additional constraint for the daughter wave can never be fulfilled in the case of plane waves \cite{LH1962}, it may hold for waves in a cylindrical container.

This has strong consequences regarding the efficiency of energy transfers through nonlinear resonant interactions. For three-wave interactions, the steepness of the daughter wave is reported in eq. \eqref{DW} and scales as $\epsilon_C \sim \epsilon_A \epsilon_B \omega_C t$. A characteristic time-scale for this nonlinear interaction, $T_{3w}$, can be obtained by assuming $\epsilon_A = \epsilon_B \equiv \epsilon$ and defining $T_{3w}$ as the solution of $\epsilon_C(T_{3w}) =\epsilon$. We obtain $T_{3w} \sim 1/(\epsilon \omega_C)$. Conversely, if three-wave resonant interactions are forbidden, the analysis has to be pursued at the next order, in which four-wave resonant interactions are found to occur for plane progressive surface gravity waves. The steepness of the daughter wave $\epsilon_D$ then scales as \cite{LH1962} $ \epsilon_D \sim \epsilon_A \epsilon_B \epsilon_C \omega_D t$, and, the time-scale of this interaction is $T_{4w} \sim 1/(\epsilon^2 \omega_D)$. In the limit of weak nonlinearities, the steepnesses remain small and therefore $T_{4w} \gg T_{3w}$. This confirms that the presence of solid boundaries enhances nonlinear energy transfers.

\textit{Experimental study\label{experiments}.} To demonstrate the efficiency of these nonlinear energy exchanges, an experiment is carried out to isolate and characterize a single three-wave resonant interaction. The main challenge is that such exchanges of energy are, most of the time, degenerate: If an interaction associated with the resonance condition
\begin{equation}
\omega_C = \omega_A \pm \omega_B, ~~~~ m_C = m_A \pm m_B,\label{18}
\end{equation}
is observed, then, since the angular frequency does not depend on the sign of $m$, i.e., upon the fact that the wave is rotating clock-wise or counter clock-wise, the following resonance condition is also verified,
\begin{equation}
\omega_C = \omega_A \pm \omega_B, ~~~~ (-m_C) = (-m_A) \pm (-m_B).
\end{equation}
This can be seen as a consequence of the symmetry of the system with respect to any plane including the vertical axis. The only situation in which such an interaction would not be degenerate is if all the waves involved are axisymmetric, that is if $m_A = m_B = m_C =0$. Such waves are experimentally difficult to generate, and require a parametric forcing as for instance a vertical oscillation of the container (leading to the so-called Faraday instability), whereas back-and-forth oscillations of the device would force the modes $m=\pm 1$. We present below an experimental setup able to isolate a single three-wave resonant interaction with non axisymmetric waves, and we focus on the resonance of the mother waves ($m_A=1$, $n_A = 1$) and ($m_B = -1$, $n_B = 2$) with the daughter wave ($m_C = 0$, $n_C = 3$).

\begin{figure}[htb]
\begin{center} \includegraphics[scale=0.65]{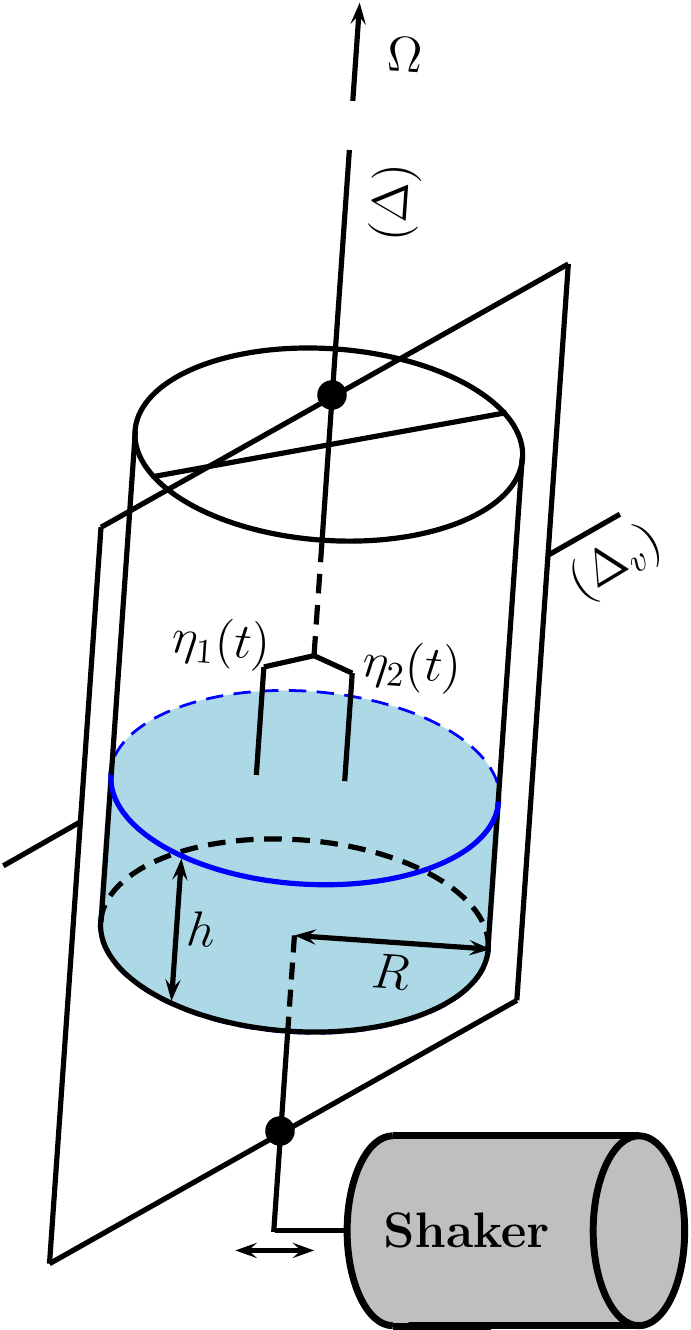}   \hspace{2cm}
\includegraphics[scale=0.20]{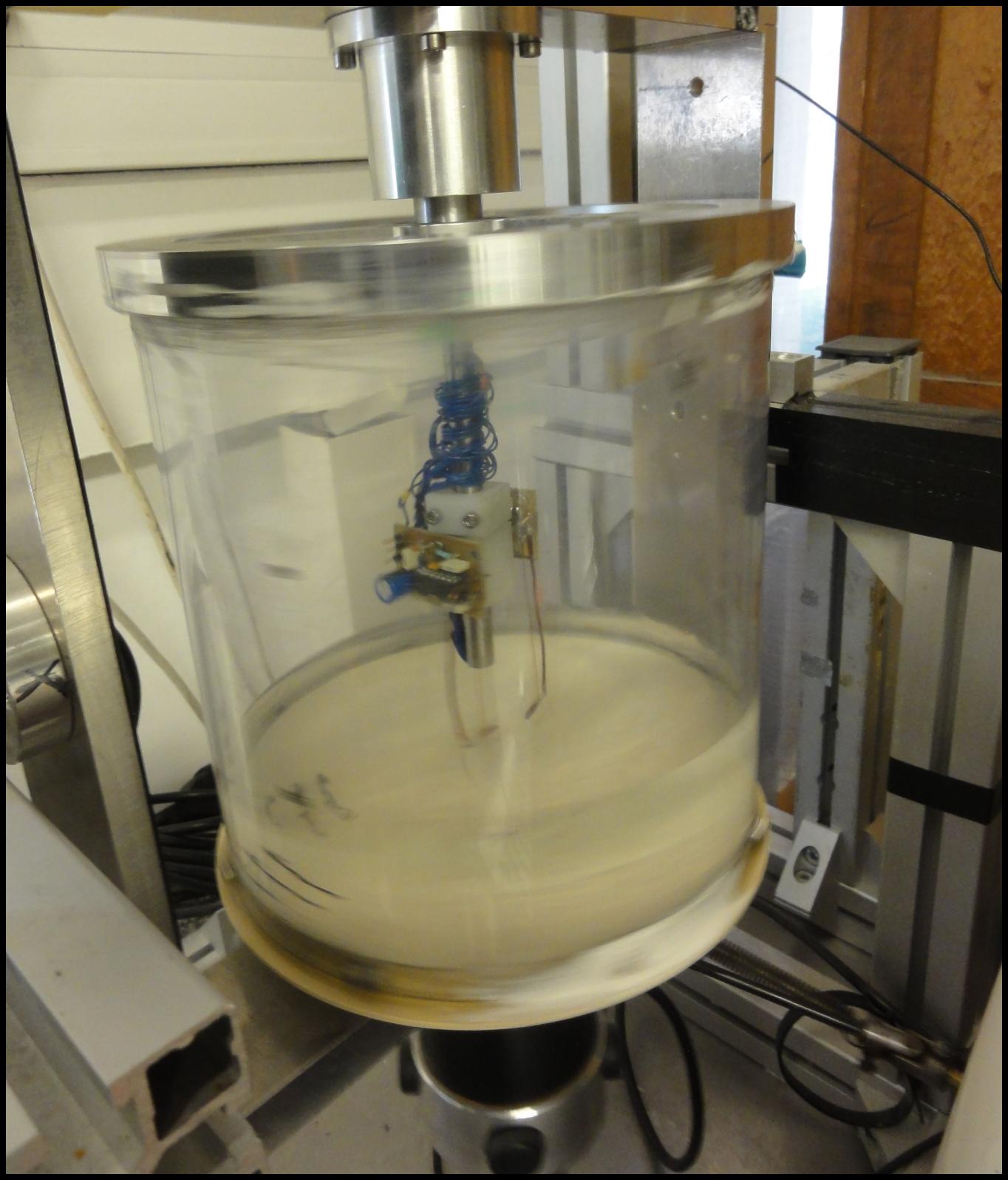}   \end{center}
    \caption{Experimental setup (sketch and photography). The notations $(r,\theta , z)$ in the text correspond to the usual cylindrical coordinate with $z=0$ at the interface. In this experiment, $h \simeq 3~\mathrm{cm}$ and $R \simeq 9~\mathrm{cm}$, while a typical value of $\Omega$ is $2\pi \times 250~ \mathrm{mHz}$.}
	\label{setup}
\end{figure}

The setup consists in a cylinder of diameter $2R=18.9~\mathrm{cm}$ filled with water up to a height $h\simeq 3~\mathrm{cm}$ such that $\pi R^2 h = 950~\mathrm{mL}$. This finite depth induces a small correction to the angular frequency of the mode $A$ ($\tanh(k_Ah) \simeq 0.6$) necessary to verify both resonance conditions \eqref{18}, the frequency mismatch $\omega_C - (\omega_A+\omega_B)$ being for the present value of $R$ too large to observe this resonance in the infinite depth limit. A motor Parvex RS420 rotates this cylinder around its vertical axis $(\Delta )$ with a constant angular frequency $\Omega$: this rotation breaks the symmetry between counter-propagating waves and, as we shall see, enables us to isolate a single resonant triad. Another shaker LDS 555 drives an oscillation of this cylinder along an axis $(\Delta_v)$ perpendicular to $(\Delta)$, which generates surface waves. Finally, two homemade capacitive height sensors record the surface elevation 2 cm away from the vertical axis, at angles $\theta = 0$ and $\theta = \pi /2$. These signals $\eta_1(t)$ and $\eta_2(t)$ are either processed by a spectrum analyser HP 35670A or recorded with a NI-USB 4431 acquisition card. This entire experimental setup is sketched in Fig. \ref{setup}.

The effect of the solid rotation of the cylinder on water waves is quantified by the non-dimensional Froude number $Fr =\Omega \sqrt{R/g}$, the most notable features being that:
\begin{itemize}

\item The equilibrium free-surface on top of which the waves develop is no longer plane but parabolic at low $Fr$.
\item New waves can be observed, as inertial waves or Rossby waves.
\item The velocity fields of surface waves evolve because of the Coriolis acceleration and of the form of the free surface. The boundary layers turn to Eckman layers and the velocity fields tend to become independent on the vertical $z$ coordinate.
\end{itemize}

In the present study, rotation is only used to disentangle counter-propagating waves, for which small Froude numbers are sufficient. In this limit $Fr \ll 1$, the effects of the solid rotation on the eigenmodes reduce to a splitting of the degenerate eigenfrequencies. This has first been computed by Thomson and Rayleigh for shallow-water surface gravity waves \cite{Kelvin, Rayleigh}, and has then been generalized by Miles to arbitrary depths \cite{Miles1964}: the natural angular frequencies in the presence of rotation are given by
\begin{equation}
\omega_{m,n}(\Omega) = \omega_{m,n}(0) + \frac{m\Omega}{\xi_{m,n}^2-m^2} \left(1+ \frac{2\xi_{m,n}h}{R\sinh \left(\frac{2\xi_{m,n}h}{R}\right)} \right) + O(\omega_{m,n}(0)Fr^2), ~~~~~ \omega_{m,n}(0)^2 = \frac{g \xi_{m,n}}{R}\tanh\left(\frac{\xi_{m,n}h}{R}\right),\label{splitting_rotation}
\end{equation}
where $\omega_{m,n}(0)$ is the natural angular frequency in the abscence of rotation. This splitting is similar to the one of acoustic modes in rotating stars (used in helioseismology, see, e.g.,  \cite{Cowling1949}) and to the Zeeman effect in optics. The effect of surface tension is discussed in \cite{Bauer1997} and large Froude numbers in \cite{Mougel2015}. According to \eqref{splitting_rotation}, the resonance condition $\omega_A + \omega_B = \omega_C$ for the parameters $\{ m_i,n_i \}$ mentioned above occurs for a rotation frequency of $242~\mathrm{mHz}$, associated to a Froude number of 0.25.

\begin{figure}[htb]
\begin{center} \includegraphics[scale=0.70]{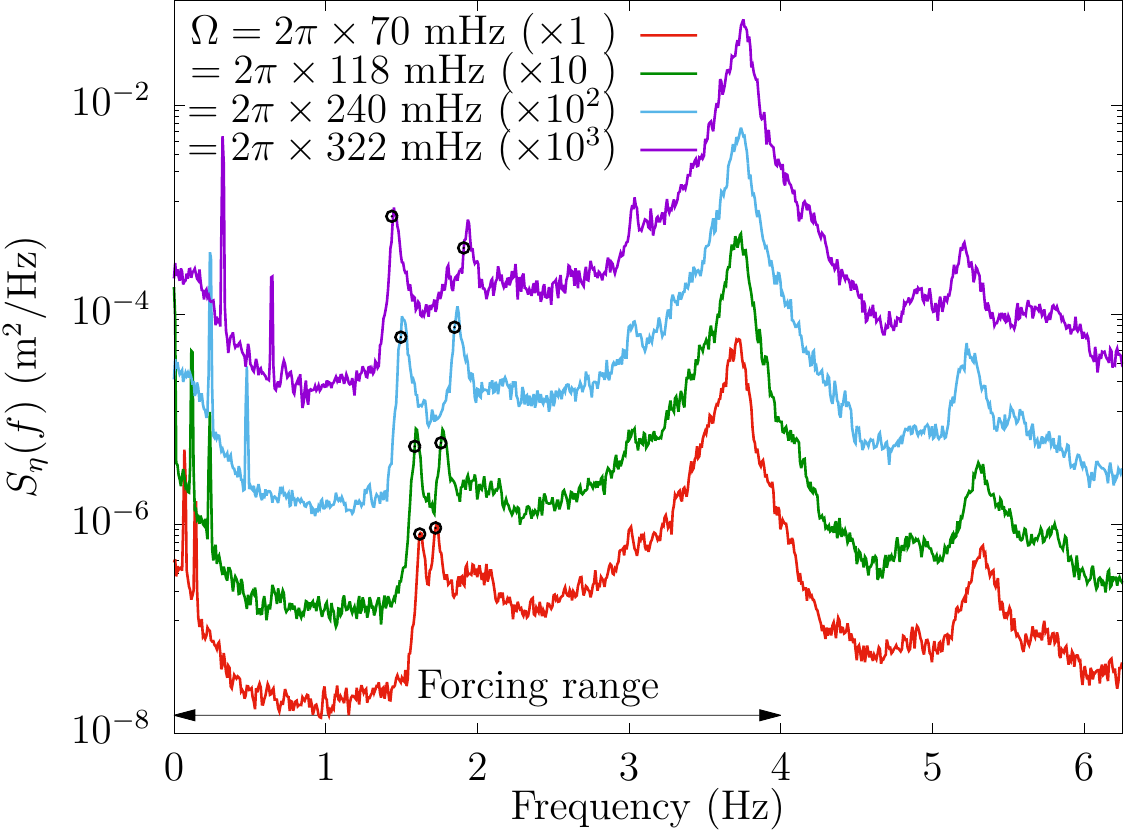}  ~~~~~\includegraphics[scale=0.70]{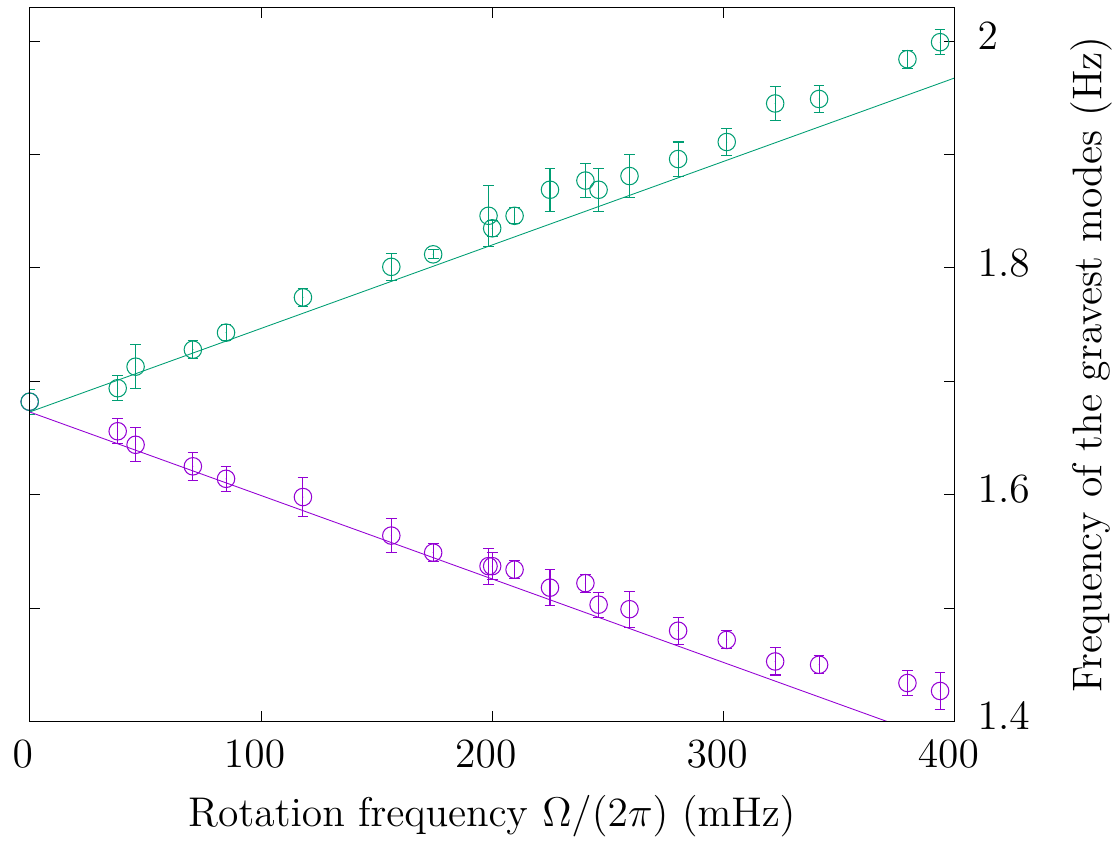}  \end{center}
    \caption{Left: Surface elevation power spectrum. The black circles are the theoretical resonance frequencies (see right graph). Right: Experimental natural frequencies of the gravest modes, compared to the theoretical prediction of \eqref{splitting_rotation} derived in the limit of small Froude numbers}
	\label{spectres}
\end{figure}

The first experiment we conduct consists in driving  the shaker with a noise of bandwidth $[ 0, 4~\mathrm{Hz}]$, for several rotation rates $\Omega$. Power spectra of the surface elevation are reported in Fig. \ref{spectres} left, and reveal three major contributions in the forcing range: two peaks around 1.8 Hz evidencing the splitting of the gravest eigenmode ($m=\pm 1$, $n=1$), and a third one around 3.8 Hz corresponding to the modes ($m=\pm 1$, $n=2$). A signal of angular frequency $\Omega$ and a harmonic at $2\Omega $ are also measured (see Fig. \ref{spectres}) as a result of mechanical imperfections during the rotation. Finally, a component of frequency close to 5.3 Hz is observed and cannot be ascribed to the forcing: it corresponds to the daughter wave $m_C = 0$ and $n_C = 3$, as shall be shown from three of its characteristics.

From these spectra, the theoretical result \eqref{splitting_rotation} of Miles \cite{Miles1964} is found to accurately describe our experiment without any fitting parameter (see Fig. \ref{spectres} right). The splitting of the other modes of frequency close to 3.8 Hz is proportional to $1/(\xi_{\pm 1,2}^2-1^2) \simeq 0.04$ (see eq. \eqref{splitting_rotation}) and cannot be resolved with this setup. From the spectrum reported in Fig. \ref{spectres} left, it is visible that the frequency of the peak out of the forcing range slightly decreases as the rotation rate increases: We checked that its angular frequency matches $\omega_{+1,1}+\omega_{-1,2}$, i.e., is related to the gravest peak. This first result confirms that the resonance condition in terms of frequencies holds, i.e., that $\omega_C= \omega_A + \omega_B$ in \eqref{18}.

\begin{figure}[htb]
\begin{center} \includegraphics[scale=0.75]{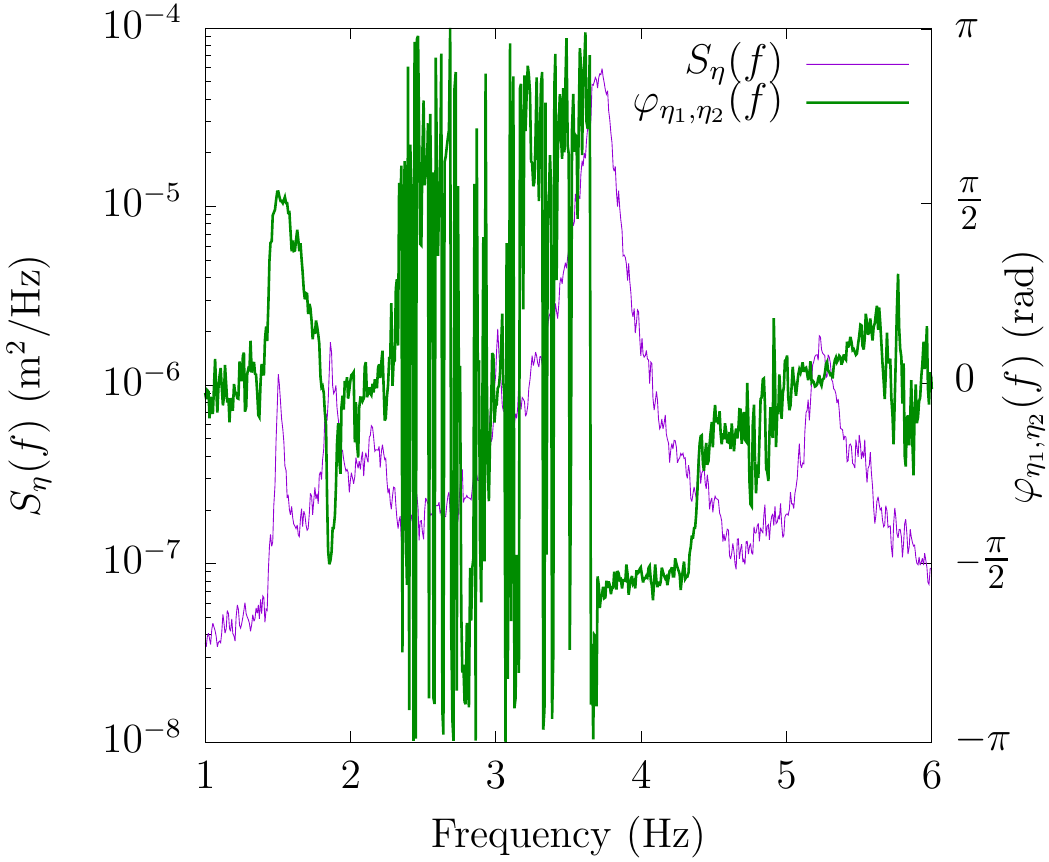} ~~~~~~~ \includegraphics[scale=0.75]{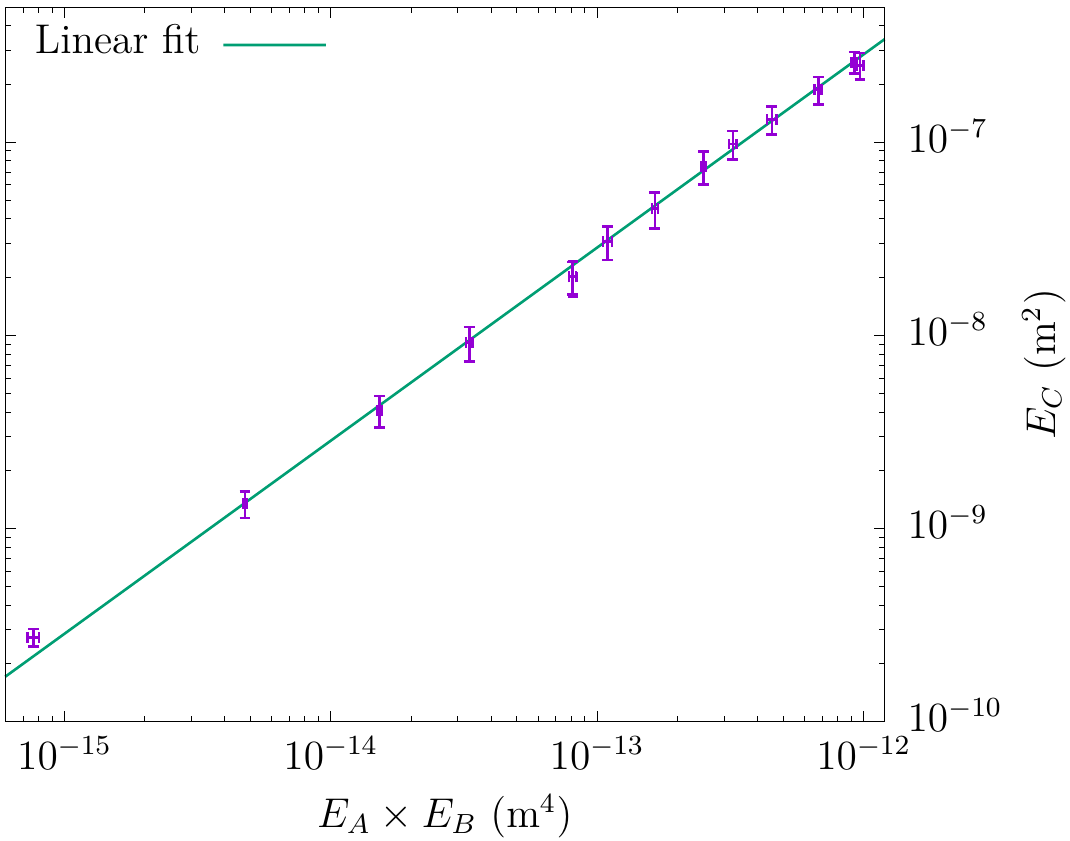}    \end{center}
    \caption{Left: surface elevation spectra and phase of the cross-spectral density for $\Omega = 2\pi \times 0.240~\mathrm{mHz}$. Right: Evolution of a measure of the energy of the daughter wave with the ones of the mother waves.}
	\label{scorr}
\end{figure}

In a second experiment, the rotation frequency is fixed at 240 mHz, with the same forcing as above. The simultaneous acquisition of the surface elevation at two different positions through the signals $\eta_1(t)$ and $\eta_2(t)$ is now used. More precisely, the phase of the cross-spectral density is computed, defined as
\begin{equation}
\varphi_{\eta_1,\eta_2}(f) = \mathrm{Angle}\left( \iint_{-\infty}^\infty \eta_1(t) \eta_2(t+\tau) e^{2i\pi f\tau} \mathrm{d}t \mathrm{d}\tau \right).
\end{equation}
When the signals $\eta_1$ and $\eta_2$ are correlated for a given frequency $f$, $\varphi_{\eta_1,\eta_2}(f)$ converges toward the phase shift between them. For instance, taking $\eta_1 \propto \mathrm{cos}(2\pi ft)$ and $\eta_2 \propto \mathrm{cos}(2\pi f t - \varphi_0)$, leads to $\varphi_{\eta_1,\eta_2}(f) = \varphi_0$. In the present experiment, the two signals are correlated when the surface elevation corresponds to an eigenmode of the form $\eta \propto \mathrm{cos} (2 \pi ft - m \theta )$. Since $\eta_1(t) = \eta(r_0, \theta = 0, t)$ and $\eta_2(t) = \eta(r_0, \theta = \pi/2, t)$ with $r_0 = 2~\mathrm{cm}$, we obtain $\varphi_{\eta_1,\eta_2}(f) = m \pi /2$. Thus, the measurement of $\varphi$ gives access to the angular index $m$. The results obtained from the experiment are reported in Fig. \ref{scorr} left. Consistently with the previous results, the couple of peaks around 1.7 Hz are found to be associated to $m = \pm 1$, the gravest one being $m_A = 1$. Moreover, the wave field at $\omega_B \simeq 2\pi \times  3.7 ~\mathrm{Hz}$ is dominated by the component $m_B = -1$. The main result is that the daughter wave is indeed a mode of angular index $m_C =0$. This also confirms that such mode cannot be directly excited through back and forced oscillation of the cylinder, and therefore results from nonlinear interactions. This verifies the second resonant condition $m_C = m_A + m_B$ in \eqref{18}.

In a third experiment, the rotation frequency is fixed, as before, at 240 mHz, and the gain of the amplifier driving the shaker is varied. Since the spectrum consists in well-defined peaks, we can evaluate a quantity $E_i$ proportional to the energy of each mode by integrating the power-spectra over a fixed and constant frequency window, 
\begin{equation}
E_i = \int_{f_i-\Delta f}^{f_i+\Delta f}S_\eta(f)  \mathrm{d}f.
\end{equation} Note that although these quantities $E_i$ slightly depend upon the choice of $\Delta f$, their relative variations considered here do not. In Fig. \ref{scorr} right, we show that the relation $E_C \propto E_A \times E_B$ is verified over four decades. This is consistent with the quadratic energy flux from the mother waves to the daughter one of eq. \eqref{DW}.

Therefore, the experiments reported here confirm the characteristics of these three-wave interactions between gravity waves in a cylinder: both resonance conditions are observed ($\omega_C = \omega_A + \omega_B$ and $m_C = m_A + m_B$), along with a quadratic relation of the form $\eta_C \propto \eta_A \times \eta_B$. Note that the theoretical work performed in the appendix does not model exactly these experiments, because the depth is not infinite, the surface tension is nonzero and rotation affects the dynamics. However these corrections remain small. Indeed, the finite-height only slighly affects the gravest mode: the non-dimensional number $\mathrm{tanh}(k h)$ introduced in \eqref{splitting_rotation} quantifies this effect and is 0.6 for the wave A, and 0.98 or more for the modes B and C. The largest frequency considered here, 5.3 Hz, also is small compared to the one at which the transition from gravity waves to capillary waves occurs ($f_\mathrm{g.c.} = 1/(2\pi) (4\rho g^3/\sigma)^{1/4} = 13~\mathrm{Hz}$ in water, where $g=9.81~\mathrm{m}\cdot \mathrm{s}^{-2}$ is the acceleration of gravity, $\sigma = 75~\mathrm{mN}\cdot \mathrm{m}^{-1}$ is the surface tension and $\rho = 10^3~\mathrm{kg}\cdot \mathrm{m}^{-3}$ is the density of water) : since no high frequency parasitic capillary wave is observed, as evident from the absence of high-frequency peak in the surface elevation power spectra, surface tension can be consistently neglected. Regarding the solid rotation, the Froude number at resonance $Fr=0.25$ remains small compared to unity. Note that dissipation would have to be taken into account to describe these steady-states, in particular for the last set of experiments. This is usually done by adding a linear damping term in the evolution equations of each mode, (see, e.g., \cite{Miles1984a,Miles1984b}). However, such linear damping coefficients have to be measured experimentally and are very sensitive to the pollution of the free surface, especially for water. In practice, nonlinear damping has to be considered when experiments are performed in a small container as a result of the walls. In addition to the nonlinear energy exchanges, this makes the quantitative study of these steady-states very challenging.

\textit{Conclusion.} In this Rapid Communication, resonant interaction between gravity waves in a circular container are characterized theoretically and experimentally. Unlike plane waves in the ocean, which undergo four-wave or higher interactions, this geometry allows for three-wave resonant interactions. This effect traces back to the resonance conditions, in which the conservation of linear pseudo-momentum is replaced by the conservation of angular pseudo-momentum. The resonance conditions and the linear growth rate are derived. An experiment able to isolate a single resonant triad, achieved by adding a small rotation to the container, is then performed to characterize the resonance. Both the resonant condition in term of angular frequency $\omega$ and of angular index $m$ are observed, along with the quadratic nature of these energy exchanges.

Such resonant interactions are of crucial importance for, at least, two separate outlooks. The first one deals with sloshing control, that aims at a reduction of the energy of the gravest eigenmode (denoted as wave A here). In this perspective, deliberately forcing another eigenmode (wave B) would result in a transfer of energy toward wave C and therefore act as an efficient damping mechanism for the mother waves. The study of such energy exchanges in a steady state, with dissipation and presumably out of the weakly nonlinear limit, would be required to fully discuss this point. A second motivating follow-up concerns wave turbulence, which predicts the statistical properties of a chaotic state in which many plane waves of small amplitudes, different frequencies and different directions interact with one another \cite{Nazarenko}. The results of this theory strongly depend on the minimal number of waves involved in a resonant interaction, and the statistics of the gravest modes in the case of a cylindrical vessel, for instance measured experimentally in \cite{Deike2012, Deike2014}, are still an open question that would require to take into account the three-wave interactions reported in the present study. This remark is not restricted to gravity surface waves. The minimal number of planes waves involved in a resonant interaction can be determined from the dispersion relation (see, \textit{e.g.}, \cite{Nazarenko}): for instance, capillary surface waves, hydroelastic waves and internal waves in stratified and/or rotating fluids can undergo resonant three-wave interactions, whereas elastic waves in thin plates and deep-water gravity waves cannot. Therefore, depending on the geometry of the experimental apparatus, the gravest modes in all of these examples may not be accurately described by plane waves and this minimal number of waves may vary, as shown here for gravity surface waves in a cylindrical vessel.

\textit{Acknowledgement.} The author is thankful to S. Fauve and F. P\'etr\'elis for fruitful discussions. This work is supported by CNES and Grant No. ANR-17-CE30-0004.

\section*{Appendix}
We consider a cylinder of radius $R$ filled with an infinite depth of water: the usual $(r,\theta, z)$ cylindrical coordinate system is used, $z=0$ corresponding to the interface between water and air. In the potential theory, the surface elevation $\eta_i$ [equation \eqref{eta_i}] of the mother waves corresponds to a velocity field
\begin{equation}
\vec{u}_i = - \vec{\nabla} \left( \Psi_i J_{m_i} (k_i r) e^{k_i z} \mathrm{sin}(\omega_i  t - m_i \theta)  \right)= - \vec{\nabla} \phi_i,\label{eigenmode}
\end{equation}
with $ \epsilon_i = \Psi_i k_i^2 / \omega_i$ \cite{Ibrahim}.
As derived by Longuet-Higgins \cite{LH1962}, the leading order correction describing the nonlinear coupling between these two waves $i=\{A,B\}$, $\vec{u}_C = - \vec{\nabla} \phi_C$, is the solution of
\begin{equation}
\partial_{tt} \phi_C + g \partial_z \phi_C \underset{z = 0}{=} 2 \partial_t \left( \vec{u}_A \cdot \vec{u}_B \right), ~~~~~ \partial_r \phi_C \underset{r=R}{=} 0,~~~~~ \triangle \phi_C = 0.\label{system}
\end{equation}
Using recurrence relations of Bessel functions, we obtain
\begin{align}\label{4}
\left( \vec{u}_A \cdot \vec{u}_B \right) \underset{z = 0}{=} \frac{\psi_A \psi_B k_A k_B}{4}[ \left( 2 J_{m_A} J_{m_B} + J_{m_A+1} J_{m_B+1} + J_{m_A-1} J_{m_B-1} \right) \mathrm{cos}(\varphi_A - \varphi_B) + \\ \left( -2 J_{m_A} J_{m_B} + J_{m_A+1} J_{m_B-1} + J_{m_A-1} J_{m_B+1} \right) \mathrm{cos}(\varphi_A + \varphi_B) ], \nonumber
\end{align}
where $\varphi_i = \omega_i t - m_i \theta$ is the total phase of the wave $i$. The remainder of the analysis consists in determining whether this term resonantly forces a free wave in \eqref{system}. To do so, equation \eqref{4} is expanded in Dini series, 
\begin{equation}
\left( \vec{u}_A \cdot \vec{u}_B \right) \underset{z = 0}{=} \sum_n \alpha_n J_{m_A - m_B} (\xi_{n,m_A-m_B} \frac{r}{R} )\mathrm{cos}(\varphi_A - \varphi_B) + \sum_n \beta_n J_{m_A + m_B} (\xi_{n,m_A+m_B} \frac{r}{R} ) \mathrm{cos}(\varphi_A + \varphi_B),\label{eq6}
\end{equation}
where $\xi_{n,m}$ is the $n$th root of $J_m'$, and the coefficients $\{ \alpha_n, \beta_n \}$ result from orthogonality relations. In particular,
\begin{equation} 
\beta_n = \frac{\Psi_A \Psi_B k_A k_B}{4} \kappa, ~~~~~ \kappa = \frac{\int_0^1 x J_{m_C}(\xi_{n,m_C}x) \left[-2 J_{m_A} J_{m_B} + J_{m_A+1} J_{m_B-1} + J_{m_A-1} J_{m_B+1} \right] \mathrm{d}x}{\int_0^1 x J_{m_C}^2(\xi_{n,m_C}x)\mathrm{d}x},\label{kappa}
\end{equation}
where $J_{m_i}$ is a shorthand for $J_{m_i}(\xi_{n_i,m_i}x)$. Since \eqref{system} is a linear equation we consider a single term in \eqref{eq6}, of the form $\beta_n J_{m_C}(k_C) \mathrm{cos}\left[ \omega_C t - m_C \theta \right]$,
with $m_C = m_A + m_B$, $\omega_C = \omega_A + \omega_B$ and $k_C = \xi_{n,m_A+m_B}/R$. 
The velocity potential forced by this term is a solution of \eqref{system} with $2 \partial_t \left( \vec{u}_A \cdot \vec{u}_B \right) = - 2\omega_C \beta_n J_{m_C}(k_C r) \mathrm{sin}\left[ \omega_C t - m_C \theta \right]$.
This system has two very different sets of solutions: a surface wave of constant amplitude that does not follow the linear dispersion relation (a ``bound wave'') if $\omega_C^2  \neq g k_C$, and a linearly growing ``free wave'' if $\omega_C^2 = g k_C$, of the form
\begin{equation}
\phi_C = \beta_n t J_{m_C}(k_C r)  \mathrm{cos}\left[ \omega_C t - m_C \theta \right] e^{k_Cz} \Rightarrow \eta_C = - \frac{\beta_n k_C t}{\omega_C} J_{m_C}(k_C r)  \mathrm{sin}\left[ \omega_C t - m_C \theta \right]. \label{deriv_DW}
\end{equation}

%
%

Equations \eqref{kappa} and \eqref{deriv_DW} lead to the result \eqref{DW}. Finally, note that \eqref{DW} evidences a phase locking between the waves: with both mother waves of the form $\mathrm{cos}(\omega_i t - m_i \theta)$, the resulting daughter wave is proportional to $\mathrm{sin}(\omega_i t - m_i \theta)$.

\end{document}